\documentclass[12pt]{article}
\usepackage{graphicx,color,pictex,fancybox,epic,epsfig,amssymb}
\begin{document}

\begin{flushright}
{\tt hep-th/0504130}
\end{flushright}

\vspace{5mm}

\begin{center}
{{{\Large \bf BPS D-branes from an Unstable
D-brane\\
in a Curved Background}}\\[14mm]
{Chanju Kim}\\[3mm]
{\it Department of Physics, Ewha Womans University,
Seoul 120-750, Korea}\\
{\tt cjkim@ewha.ac.kr}\\[7mm]
{Yoonbai Kim,~~Hwang-hyun Kwon,~~O-Kab Kwon}\\[3mm]
{\it BK21 Physics Research Division and Institute of
Basic Science,\\
Sungkyunkwan University, Suwon 440-746, Korea}\\
{\tt yoonbai@skku.edu~~hhkwon@skku.edu~~okab@skku.edu}
}
\end{center}
\vspace{10mm}

\begin{abstract}
We find exact tachyon kink solutions of DBI type effective action
describing
an unstable D5-brane with worldvolume gauge field turned on in
a curved background. The background of interest
is the ten-dimensional lift of the Salam-Sezgin vacuum and, in the
asymptotic limit, it approaches
${\rm R}^{1,4}\times {\rm T}^2\times {\rm S}^3$.
The solutions are identified as
composites of lower-dimensional D-branes and fundamental strings,
and, in the BPS limit, they become a D4D2F1 composite wrapped on
${\rm R}^{1,2}\times {\rm T}^2$ where ${\rm T}^2$ is inside
${\rm S}^3$.
In one class of
solutions we find an infinite degeneracy with respect to a constant
magnetic field along the direction of NS-NS field on ${\rm S}^3$.
\end{abstract}

\newpage

\setcounter{equation}{0}
\section{Introduction}

Study of unstable D-branes in string theory has led to a deeper
understanding of the theory in various aspects~\cite{Sen:2004nf}.
In a dynamical aspect, it provides an example of time-dependent
string background that can be described by the language of
worldsheet conformal field theory~\cite{Sen:2002nu}.
Descent relations among BPS and non-BPS D-branes of various
dimensions provided a new perspective on the characteristics of
D-branes and hepled the development of the classification of
D-brane charges in terms of K-theory~\cite{Witten:1998cd}.
These rolling tachyons and tachyon solitons can also be dealt,
at least qualitatively, in terms of DBI type effective
action~\cite{Sen:2002an,Sen:2003tm}.

Most of the studies so far have been performed on flat unstable
D-branes. The purpose of the paper is an attempt to extend the analyis of
unstable D-branes to a curved bulk background and find tachyonic
kink solutions on them. The background that we will consider is
the ten-dimensional embedding of the supersymmetric vacuum, ${\rm
R}^{1,3}\times {\rm S}^2 $, of the Salam-Sezgin
model~\cite{Cvetic:2003xr}. The model is a gauged ${\cal N}=(1,0)$
supergravity in six dimensions coupled to a tensor and an abelian
vector multiplet~\cite{Salam:1984cj}. The vacuum solution is the
unique non-singular solution of the model with maximal
four-dimensional spacetime symmetry. It can be embedded into ten
dimensions as the type IIA supergravity background solution, ${\rm
R}^{1,3}\times {\rm S}^2 \times {\cal H}^{2,2}\times {\rm S}^1$,
where ${\cal H}^{2,2}$ represents a three-dimensional hyperboloid.
In view of recent works on the dynamics of D-branes in  NS5-brane
background initiated by D. Kutasov~\cite{Kutasov:2004dj}, it is interesting to notice
that~\cite{Cvetic:2003xr}, in the asymptotic limit at large
distances, the local geometry of the background of Salam-Sezgin
vacuum approaches the NS5-brane near-horizon
geometry~\cite{Callan:1991at,Rey:1989xj}. However, there is a difference in
that string coupling constant goes to zero in the asymptotic limit
of the background, while it blows up in the throat region of the
NS5-brane. In this limit, it is valid to study non-BPS D-branes in
terms of DBI type effective theory. Still our work may have an
implication for recent developments along the line
of~\cite{Kutasov:2004dj}.

In the asymptotic limit, the background approaches
${\rm R}^{1,4}\times {\rm T}^2\times {\rm S}^3$.
The object of consideration is a non-BPS D5-brane whose
worldvolume lies on ${\rm R}^{1,2}\times {\rm S}^{3}$.
After the unstable D-brane decays, one may expect the
generation of lower-dimensional stable
brane configurations~\cite{Sen:2003tm,Kim:2003in}
and emission of energy to closed string degrees of
freedom~\cite{Okuda:2002yd,Lambert:2003zr}.
The former, specifically codimension-one object, is what we would like to
study in this paper. We find two classes of exact solutions both of which
are identified as thick D4-branes on ${\rm R}^{1,2}\times {\rm T}^{2}$
where the two-torus is embedded in the three-sphere.
One class of the solutions contains fundamental string and two D2-branes;
one is tubular and the other flat. The flat D2-brane disappears
when a constant magnetic field, $h$, along the direction of NS-NS
field on S$^3$ vanishes.
In the thin limit, the solution
becomes BPS when $h=0$ and the energy expression is given by a
BPS sum rule.
The other class involves one tubular D2-brane and fundamental
string.
In the thin limit it becomes essentially identical to
the former class of the solutions except that it has infinite degeneracy
with respect to $h$.

The rest of this paper is organized as follows. In section 2,
we give a brief review of the bulk background on which we consider
tachyon kink solutions of an unstable D5-brane.
In section 3, by considering an unstable D5-brane in the given background,
we obtain stable codimension-one D-brane configurations and identify
their BPS limit. We conclude the paper in section 4.

\setcounter{equation}{0}
\section{Background Geometry}

In this section we briefly describe the background of the ten-dimensional
lift of Salam-Sezgin vacuum\footnote{The details of the construction 
of the background can be found in~\cite{Cvetic:2003xr}.}
and set the  convention. The type IIA
supergravity action is given  by
\begin{equation}\label{nsaction}
S_{\rm NS} = {1\over 2\kappa^2_{10}} \int d^{10}x \sqrt{-G}\;
e^{-2\Phi} \Big( R + 4\partial_\mu \Phi \partial^\mu \Phi
-{1\over 2\cdot 3!} H^{\mu\nu\rho} H_{\mu\nu\rho} \Big)
\end{equation}
for the massless NS-NS fields in the string frame.
Here
$\kappa_{10}$ is the ten-dimensional gravitational coupling
constant, $\Phi$ is the dilaton field, and $H_{\mu\nu\rho}$ is
the field strength of NS-NS 2-form potential $B_{\mu\nu}$. R-R
and fermionic fields do not play a role for the embedding and
may be set  to  zero.

The six-dimensional Salam-Sezgin model is a gauged ${\cal
N}=(1,0)$ supergravity coupled to a tensor and an abelian
vector multiplet. Its bosonic sector consists of the
metric, real scalar field, $\phi$, abelian one-form and
two-form gauge fields, $A$ and $B'$.  Through a chain of
dimensional reductions and truncations, one can see that the
model can be obtained by dimensionally reducing IIA supergravity on
${\cal H}^{2,2}\times {\rm S}^1$. Here
${\cal H}^{2,2}$ is the quadric, $\mu_1^2 + \mu_2^2 - \mu_3^2
-\mu_4^2 =1, $ in Euclidean space ${\rm R}^4$. The embedding is
described by the ansatz
\begin{eqnarray}\label{ansatz}
ds^2 &=& e^{-\phi/2} ds_6^2 + {1\over 2g^2} \left[
d\rho^{2}+\frac{\cosh^{2}\rho}{\cosh 2\rho}(d\sigma
-g A )^{2}
+\frac{\sinh^{2}\rho}{\cosh 2\rho}(d\tau + g A)^{2} \right]
+d\chi^2 \, ,\nonumber \\
H &=& \frac{\sinh\rho\cosh\rho}{g^2(\cosh 2\rho)^{2}} \;
d\rho\wedge (d\sigma-gA)\wedge (d\tau +gA)\nonumber\\
&&+\frac{1}{2g \cosh 2\rho}\; {\rm d} A
\wedge \left[\cosh^{2}\rho(d\sigma-gA)-
\sinh^{2}\rho(d\tau +gA)\right] + {\rm d} B', \\
e^\Phi &=& (\cosh 2\rho)^{-1/2} \, e^{-\phi/2}, \nonumber
\end{eqnarray}
where $g$ is a rescaled coupling constant for
the gauge potential $A$, $(\rho,\sigma,\tau)$ parametrize the
hyperboloid ${\cal H}^{2,2}$ and $\chi$ the circle ${\rm S}^1$.

Using the ansatz (\ref{ansatz}), any solution of the Salam-Sezgin
model  can be lifted to ten dimensions. The one of our interest
is the supersymmetric vacuum solution, ${\rm R}^{1,3}\times {\rm
S}^2$,
with the magnetic monopole flux on ${\rm S}^2$, which is given by
\begin{eqnarray}\label{ssvacuum}
ds_6^2 &=& dx_4^2 + {1\over 8g^2} (d\theta^2 + \sin^2\theta
d\varphi^2) \, , \nonumber \\
A &=& -{1\over 2g} \cos\theta d \varphi, \quad {\rm d} B' =0,
\quad \phi =0
\end{eqnarray}
where $dx_4^2$ represents the line element of the
four-dimensional Minkowski space. Inserting the solution
(\ref{ssvacuum}) into the ansatz (\ref{ansatz}) gives the
background of the ten-dimensional lift of Salam-Sezgin vacuum.
In the large-$\rho$ limit, the embedded
solution simplifies and can be written as
\begin{eqnarray}\label{asymptotic}
ds^2 &=& dx_6^2 + {1\over 2g^2} d\rho^2 +{1\over 8g^2}
\left[ d\theta^2 + \sin^2\theta \, d\varphi^2 + \left(
d(\sigma-\tau) +\cos\theta \, d\varphi \right)^2 \right], \nonumber \\
H &=&  {1\over 8g^2} \sin\theta \, d\theta \wedge d \varphi
\wedge d (\sigma-\tau), \quad \Phi = -\rho
\end{eqnarray}
where  $ 0\leq
\theta \leq \pi,\, 0\leq \varphi \leq 2\pi,\, 0\leq (\sigma
-\tau) \leq 4\pi $ and $dx_6^2$ represents the line element of ${\rm
R}^{1,3}\times {\rm T}^2 $. Besides the fact that there is a
linear dilaton background along the direction ${\rm R}_\rho$,
one can notice that the coordinates $(\theta, \varphi ,
\sigma-\tau)$ parametrize ${\rm S}^3$ as the Hopf fibration of ${\rm S}^1$
over ${\rm S}^2$ and the field  strength $H$ is proportional to the
volume form of the unit three sphere. Therefore,  the background
(\ref{asymptotic}) is locally identical to the near-horizon
geometry  of the NS5-brane~\cite{Callan:1991at,Rey:1989xj}.  However, as
remarked in the previous section,  the behavior of the dilaton is
opposite in the respective limits of the two background
solutions.

In the following discussions, it
turns out to be useful to rescale the angular variables
by the radius of ${\rm S}^3$, $R= 1/\sqrt{8g^2} $, to the variables
$(u,v,w)$ and use a gauge-fixed value of the gauge potential
$B$, which brings (\ref{asymptotic}) to
\begin{eqnarray}
ds^{2}&=& dx_6^{2} +4R^{2}d\rho^{2} +
du^2 + \sin^2 \left(\frac{u}{R}\right)\, dv^2 +\left[dw + \cos
\left(\frac{u}{R}\right)\, dv\right]^{2},
\label{bgS3}\\
B &=& -\cos \left(\frac uR \right) dv \wedge dw ,
\label{NSNS} \\
\Phi &=& -\rho .\label{dagain}
\end{eqnarray}

\setcounter{equation}{0}
\section{D4D2F1-composites from Unstable D5-brane}

In this section we will study tachyon kink solutions on an
unstable D5-brane in the large-$\rho$ limit of the ten-dimensional lift of
Salam-Sezgin vacuum on 
${\rm R}^{1,3}\times {\rm T}^2\times {\rm R}_\rho \times {\rm S}^3$ 
described by Eqs.~(\ref{bgS3}), (\ref{NSNS}), and (\ref{dagain}).
We consider the D5-brane on R$^2\times$S$^3$ with the
coordinates  $(z, \rho, u , v, w)$ where $z$ is one of the spatial
coordinates\footnote{Inclusion of the linear dilaton 
coordinate $\rho$ in the worldvolume of D5 is not essential for the
type of solutions we consider; it can be
trivially replaced by any other spatial coordinate on ${\rm R}^{1,3}$
without changing the rest of the analysis.}
of ${\rm R}^{1,3}$.
String coupling constant goes to
zero in the limit and many features of the dynamics of unstable
D-branes can
be described by the DBI-type worldvolume effective
action~\cite{Garousi:2000tr}
\begin{equation}\label{d9a}
S_{{\rm D}5}
= -{\cal T}_{5} \int d^{6}\xi \; e^{-\Phi} V(T)
  \sqrt{-\det (g_{\mu\nu} + F_{\mu\nu} +
B_{\mu\nu}+\partial_\mu T\partial_\nu T )}\, ,
\end{equation}
where ${\cal T}_{5}$ is the tension of the non-BPS D$5$ brane,
$F_{\mu\nu}$ is the field strength of the U(1) worldvolume gauge
field,
$T(x)$ is the real tachyon field, and $\xi^\mu (\mu =
0,\ldots,5)$ runs over the coordinates $(t, z, \rho, u , v, w)$.
The background metric $g_{\mu\nu}$ and NS-NS field $B_{\mu\nu}$ are
given by the pullbacks of (\ref{bgS3}) and (\ref{NSNS}) on the D5-brane.
For the tachyon potential $V(T)$, any runaway potential with $V(0)=1$ and
$V(\pm\infty)=0$ is allowed
for the existence of D-brane configuration of our interest that
is consistent with
universal behavior in tachyon condensation~\cite{Sen:2003tm}, but
we assume a specific form for exact solutions~\cite{Buchel:2002tj}
\begin{equation}\label{V3}
V(T)=\frac{1}{\cosh \left(\frac{T}{R}\right)},
\end{equation}
where $R$ turns out to be  identical to the
compactification scale
$R$ in (\ref{bgS3}) for the kink solutions we
find.

For the rest of the paper we will study codimension-one solutions
of (\ref{d9a}) under the ansatz
\begin{equation}\label{ant}
T=T(u,w),
\end{equation}
and the nonvanishing components of the worldvolume gauge field
strength are
\begin{equation}\label{anf}
F_{0z}=E_{z}(u,w),~F_{vz}=\alpha(u,w),
~F_{vw}=h(u,w).
\end{equation}
It turns out to be an appropriate ansatz for the tachyon and the
U(1) gauge field to support a tachyon tube embedded in a codimension-one
D4-brane\footnote{One can also consider solutions with all $w$ replaced
by $v$ in (\ref{ant}) and (\ref{anf}). Since the coordinates $v$ and $w$ are
symmetric (except minus signs in $vw$-components of two-form
fields and the range of the variables), the resulting solutions will be
essentially identical to the present case.}.
With this ansatz, the coordinate $\rho$ decouples from the
others and hence the presence of the dilaton background  (\ref{dagain})
plays no role. Therefore, in the following,
we will simply ignore the dilaton background in the action.

Applying the Bianchi identity
\begin{equation}\label{nBi}
\partial_{\mu}F_{\nu\lambda}+\partial_{\nu}F_{\lambda\mu}+\partial_{\lambda}
F_{\mu\nu}=0,
\end{equation}
we can further simplify the ansatz (\ref{anf}). The result is
\begin{equation}\label{anf2}
E_{z}={\rm constant},\quad \alpha={\rm constant},\quad h=h(w).
\end{equation}
Substituting the nonvanishing gauge field (\ref{anf2}),
the tachyon (\ref{ant}), and the NS-NS two-form field (\ref{NSNS}) into the
action (\ref{d9a}), we have
\begin{eqnarray}\label{d9b}
S_{{\rm D}5}
&=& -{\cal T}_{5} \int d^{6}\xi\; V(T)
\Bigg\{
(1-E_{z}^{2}+\alpha^{2})\left[1+(\partial_{u}T)^{2}+(\partial_{w}T)^{2}\right]
\nonumber\\
&&\hspace{16mm}
+(1-E_{z}^{2})\left[1+(\partial_{u}T)^{2}\right]\left[
\left(h-\cos\left(\frac{u}{R}\right)\right)^{2}-\cos^{2}
\left(\frac{u}{R}\right)\right]
\Bigg\}^{1/2} .
\end{eqnarray}
As has been done for every tubular object, we take critical electric field
along $z$-direction, $|E_{z}|=1$, as a basic ansatz, and then
the action (\ref{d9b}) takes a simple form,
\begin{equation}\label{acD9}
S_{{\rm D}5} = -{\cal T}_{5} \,\alpha \int d^{6} \xi\, V(T)
\sqrt{1 + (\partial_u T)^2  + (\partial_w T)^2}\, .
\end{equation}

Then the equations of motion for the tachyon and gauge
field are given by
\begin{eqnarray}
&&
\partial_{u}\left[\frac{\alpha V}{\sqrt{-X}} \partial_{u}T\right]
+\partial_{w}\left[\frac{\alpha V}{\sqrt{-X}} \partial_{w}T\right]
= \frac{\sqrt{-X}}{\alpha}\frac{dV}{dT},
\label{taeq}\\
&&
\partial_{u}\left[\frac{\alpha V}{\sqrt{-X}}h\partial_{u}T\partial_{w}T\right]
=\partial_{w}\left[\frac{\alpha V}{\sqrt{-X}}h\left(1+(\partial_{u}T)^{2}\right)
\right],
\label{gae1}\\
&&\partial_{u}\left[\frac{\alpha V}{\sqrt{-X}}\cos\left(\frac{u}{R}\right)
\partial_{u}T\partial_{w}T\right]=\partial_{w}\left[\frac{\alpha V}{\sqrt{-X}}
\cos\left(\frac{u}{R}\right)\left(1+(\partial_{u}T)^{2}\right)\right],
\label{gae2}
\end{eqnarray}
where $X=-\alpha^{2}[1 + (\partial_u T)^2  + (\partial_w T)^2]$.
It is not difficult to solve these equations.
After some
manipulations with (\ref{taeq}) and (\ref{gae2}) we find that they are
consistent only when
\begin{equation}\label{tuw}
\partial_{u}T\partial_{w}T=0,
\end{equation}
i.e., the tachyon field $T$ is a function of either $u$ or $w$ but not both.
Then from (\ref{gae1}) and (\ref{gae2}) it is easy to see that
$h$ should be a constant.

\subsection{$T=T(u)$}
We first consider the case $\partial_{w}T=0$, i.e., $T$ depends only on $u$.
Then the only nontrivial equation is (\ref{taeq}) (with $\partial_w T=0$).
This equation is actually exactly the same as that for the usual
tachyon
tube~\cite{Kim:2003uc,Kim:2004xk} in lower dimensions and
can also be directly derived from the reduced action
\begin{equation}\label{acD}
S_{{\rm D}5} = -{\cal T}_{5} \,\alpha \int d^{6} \xi\, V(T)
\sqrt{1 + \left(\frac{dT}{du}\right)^2}\, ,
\end{equation}
It is now quite straightforward to solve Eq.~(\ref{taeq}).
Rewriting the equation, we find
\begin{equation}
\partial_u\left[\frac{V(T)}{\sqrt{-X}} \right] = 0.
\end{equation}
Thus the whole equations of motion reduce to a
single first-order differential equation
\begin{equation}\label{tdx}
\beta=\frac{{\cal T}_{5}V(T)}{\sqrt{-X}},
\end{equation}
where $\beta$ is an integration constant.
With the tachyon potential (\ref{V3}), we can find the exact solution in
a closed form,
\begin{equation}\label{tub}
\sinh\left(\frac{T(u)}{R}\right)
=\pm \sqrt{\left(
\frac{{\cal T}_{5}}{\alpha\beta}\right)^{2}-1}
\,\cos \left(\frac{u}{R}\right).
\end{equation}

The energy-momentum tensor $T^{\mu\nu}$ of the system is given by
\begin{equation}\label{emte}
T^{\mu\nu}=\frac{{\cal T}_{5}V}{\sqrt{-X}\sqrt{-g}}C^{\mu\nu}_{{\rm s}},
\end{equation}
where $C^{\mu\nu}_{{\rm s}}$ is the symmetric part of the cofactor of
$X_{\mu\nu} = g_{\mu\nu} + F_{\mu\nu} +
               B_{\mu\nu}+\partial_\mu T\partial_\nu T$.
For the solution (\ref{tub}), nonvanishing components are
\begin{eqnarray} \label{10te}
T^{tt} &=& \frac1{\sin(u/R)}
\left[ 1+\alpha^2 + h^2 - 2h \cos\left(\frac{u}{R}\right) \right] \Sigma(u),
\nonumber \\
T^{zz} &=& -\frac1{\sin(u/R)}
\left[ 1 + h^2 - 2h \cos\left(\frac{u}{R}\right) \right] \Sigma(u),\nonumber\\
T^{uu} &=& -\frac{\beta \alpha^2}{\sin (u/R)}, \nonumber \\
T^{ww} &=& -\frac{\alpha^2}{\sin (u/R)} \Sigma(u), \nonumber \\
T^{tv} &=& -\frac{\alpha}{\sin (u/R)} \Sigma(u), \nonumber \\
T^{tw} &=& \alpha \cot \left(\frac{u}{R}\right) \Sigma(u), \nonumber \\
T^{zw} &=& \frac{\alpha}{\sin (u/R)}
\left[ h-\cos\left(\frac{u}{R}\right) \right] \Sigma(u),
\end{eqnarray}
where
\begin{eqnarray}
\Sigma(u) &=& \beta [ 1 + (\partial_{u}T)^{2} ]
=\frac{1}{\beta\alpha^{2}}({\cal T}_{5}V)^{2} \nonumber \\
&=& \frac{{\cal T}_5^2/\beta \alpha^2}%
{[ (\frac{{\cal T}_5}{\beta \alpha})^2 - 1 ] \cos^2(\frac uR) + 1}.
\end{eqnarray}
The electric flux $\Pi_i$, which is the conjugate momentum of $A_i$, is
calculated as
\begin{equation}
\Pi_i = \frac{{\cal T}_{5}V}{\sqrt{-X}}C_{\rm A}^{0i},
\end{equation}
where $C_{\rm A}^{0i}$ is the antisymmetric part of the cofactor of
$X_{\mu\nu}$. Then the solution (\ref{tub}) has two nonzero components,
\begin{eqnarray} \label{flux}
\Pi_{z} &=& \left[1+h^2 - 2h \cos\left(\frac{u}{R}\right)\right] \Sigma(u),
\nonumber \\
\Pi_{w} &=& -\alpha \left[h-\cos\left(\frac{u}{R}\right)\right] \Sigma(u).
\end{eqnarray}

Note that, except $T^{uu}$, all nonzero components of
$T^{\mu\nu}$ and $\Pi_i$
depend on $\Sigma(u)$ which has a peak at $u=\pi R/2$ where NS-NS two-form
background field (\ref{NSNS}) vanishes. In fact, in the limit
$\beta \rightarrow 0$ it becomes a delta function
\begin{equation} \label{thin}
\Sigma(u) \longrightarrow \frac{\pi R {\cal T}_5}\alpha
\delta(u-\pi R/2),
\end{equation}
and so do the nonvanishing components of $T^{\mu\nu}$ and $\Pi_i$
($T^{uu}$
goes to zero in this limit). Therefore,  when $\beta$ is small,
the coordinate
$u$ is essentially  fixed at $\pi R/2$ and the solution
represents a
dimensionally reduced  configuration. From the background metric
(\ref{bgS3}),
we see that the configuration  spans ${\rm T}^{2}$ in the
three-sphere:
\begin{equation} \label{dst2}
ds_{{\rm S}^{3}}=du^{2}
+ \sin^2 \left(\frac{u}{R}\right)\, dv^2 +\left[dw + \cos
\left(\frac{u}{R}\right)\, dv\right]^{2}
~\stackrel{\frac{u}{R}=\frac{\pi}{2}}{\longrightarrow}~
ds_{{\rm T}^{2}}=dv^2 + dw^2 .
\end{equation}
Of course, this is because the three-sphere with the above metric
is a Hopf fibration of S$^2$ with coordinates $(u,v)$ and hence
the circle $u=\pi R/2$ along the equator of the two-sphere  corresponds
to ${\rm T}^{2}$ in the three-sphere.

 The coupling to the bulk R-R
fields can be read off from the Wess-Zumino term for unstable
D-branes~\cite{Sen:2003tm,Okuyama:2003wm},
\begin{equation} \label{wzterm}
S_{\rm WZ} = {\cal T}_{5}\int V(T)\, dT \wedge C_{{\rm RR}}
\wedge e^{F+B}.
\end{equation}
For the solution (\ref{tub}),
\begin{eqnarray} \label{dtfb}
dT &=& \partial_u T \, du, \nonumber \\
F+B &=& dt\wedge dz + \alpha\,dz\wedge dv
+ \left[ h - \cos\left(\frac uR \right) \right] dv\wedge dw.
\end{eqnarray}
Then, in the thin limit $\beta \rightarrow 0$ we can use (\ref{thin})
to simplify $S_{\rm WZ}$,
\begin{eqnarray}
S_{\rm WZ} &=& \mp \pi R {\cal T}_5 \int \left[ C_{(5)}
+ C_{(3)} \wedge (dt\wedge dz + \alpha \,dz\wedge dv
+ h\,dv\wedge dw) \right. \nonumber\\
&&\hspace{22mm} \left. +hC_{(1)}\wedge dt\wedge dz\wedge dv\wedge dw \right].
\end{eqnarray}
where the terms containing a R-R form wedged to  $dt$ are
irrelevant. The resulting configuration consists of the following
objects. First we have a D4-brane stretched along ${\rm R}^2
\times {\rm T}^2$ with coordinates $(z,\rho,v,w)$. Its RR-charge
reads
\begin{equation}
{\cal T}_4 = \pi R {\cal T}_5,
\end{equation}
which is precisely the relation one would expect when the
codimension-one solution in the worldvolume theory of an unstable
D$p$-brane
on ${\rm R}^{2}\times {\rm S}^{3}$ is
identified as a BPS D$(p-1)$-brane on R$^2\times$T$^2$.
We also have two D2-branes with charges per unit area,
$\pi R {\cal T}_5 \alpha$ and $\pi R {\cal T}_5 h$, which are
spanned by the
worldvolume coordinates $(t,\rho,w)$ and $(t,\rho,z)$, respectively. In
addition, there are fundamental strings with flux (\ref{flux}) on
cylinder ${\rm R}\times
{\rm S}^{1}$ of $(z,w)$.

In order to study the BPS nature of the solution, we now investigate the
energy-momentum tensor (\ref{10te}). For the solution to be a BPS object,
it is required that the stress components vanish in the
transverse directions,
i.e., $T^{\mu\nu} = 0$ with $\mu,\nu = u,v$. From (\ref{10te}), this is
satisfied if $\beta \rightarrow 0$ for which the pressure in $u$-direction
$T^{uu}$ vanishes (The other limit
$\alpha \rightarrow 0$ with finite $\beta$ is not interesting since
$\Sigma(u)$ and local densities blow up.) Also, the off-diagonal stress
component between D2-branes $T^{zw}$ should vanish. In the thin limit
$\beta \rightarrow 0$ where $u$ is fixed to  $\pi R/2$, this
dictates that
$h = 0$. Then we have only one D2-brane along the directions
$(\rho, w)$ and an
electric flux along the $z$-direction since
$\Pi_w$ becomes zero. This solution
is expected to form a BPS configuration. Indeed, the energy per unit area of
coordinates $(\rho , w)$ takes the form
\begin{eqnarray} \label{bpssr}
\frac{\cal E}{\int d\rho\,dw}
&=& \int dz\,du\,dv\,(1+  \alpha^2)\Sigma(u) \nonumber \\
&=& \int dz\,du\,dv\,\Pi_z
+ \pi R {\cal T}_5 \alpha \int dz\,dv\, \nonumber \\
&=& Q_{\rm F1} + Q_{\rm D2},
\end{eqnarray}
where $Q_{\rm F1}$ is the total charge of fundamental strings
along the $z$-direction and
$Q_{\rm D2}$ the total charge of D2-brane stretched along $(\rho,
w)$-direction
on the area $\int dz\,dv$.
This is the familiar BPS sum rule which we have met in our
previous works when considering solutions such as tachyon kinks and tubes
\cite{Kim:2003in,Kim:2003uc,Kim:2004xk}.
Other nonvanishing components of the energy-momentum tensor are
$T^{tv}$ whose presence means that the configuration carries
angular momentum
as in tachyon tubes, $T^{zz}$ corresponding to  the fundamental
string charge, and
$T^{ww}$ the D2-brane charge.

In summary, when $h=0$, the solution produces a BPS
D4D2F1-composite in the thin
limit, $\beta \rightarrow 0$. It consists of
the D4-brane wrapped on R$^{2}\times$T$^2$,  the tubular
D2-brane with the coordinates $(\rho,w)$, and the fundamental
strings stretched along the $z$-direction.

\subsection{$T=T(w)$}
When we put $\partial_u T=0$ into the equations of motion
(\ref{taeq})--(\ref{gae2}),
we obtain exactly the same equations as in the previous case if
$\partial_u T$ is replaced by $\partial_w T$. The
solution can be expressed as
\begin{equation}\label{tub9}
\sinh\left(\frac{T(w)}{R}\right)
=\pm \left[\sqrt{\left(
\frac{{\cal T}_{5}}{\alpha\beta}\right)^{2}-1}
\,\cos \left(\frac{w}{R}\right)\right],
\end{equation}
where $\beta$ is again given by (\ref{tdx}) with
$X = -\alpha^2[1+(\partial_wT)^2]$.

The energy-momentum tensor (\ref{emte}), however, has different
form since the background fields, (\ref{bgS3}) and (\ref{NSNS}),
are not symmetric under the exchange of $u$ and $w$.
Its nonvanishing components are
\begin{eqnarray} \label{10tew}
T^{tt} &=& \frac{1+\alpha^2}{\sin(u/R)} \Sigma(w)
+ \frac{\beta}{\sin(u/R)} \left[h^2 - 2h \cos\left(\frac uR\right) \right],
\nonumber \\
T^{zz} &=& - \frac{1}{\sin(u/R)} \Sigma(w)
- \frac{\beta}{\sin(u/R)} \left[h^2 - 2h \cos\left(\frac uR\right) \right],
\nonumber \\
T^{uu} &=& -\frac{\alpha^2}{\sin (u/R)} \Sigma(w), \nonumber \\
T^{ww} &=& -\frac{\beta\alpha^2}{\sin (u/R)}, \nonumber \\
T^{tv} &=& -\frac{\alpha}{\sin (u/R)} \Sigma(w), \nonumber \\
T^{tw} &=& \alpha\beta \cot \left(\frac{u}{R}\right) , \nonumber \\
T^{zw} &=& \frac{\alpha\beta}{\sin (u/R)}
\left[h - \cos\left( \frac uR\right) \right].
\end{eqnarray}
We also have the electric fluxes
\begin{eqnarray} \label{fluxw}
\Pi_{z} &=& \Sigma(w)
+ \beta \left[h^2 - 2h \cos\left(\frac{u}{R}\right)\right], \nonumber \\
\Pi_{w} &=& -\alpha \beta \left[h-\cos\left(\frac{u}{R}\right)\right].
\end{eqnarray}

Comparing with the previous case, we see that there are terms which are
not proportional to $\Sigma(w)$. But note that they are all multiplied by
$\beta$ and, hence, all the quantities are proportional to
$\Sigma(w)$ in
$\beta \rightarrow 0$ limit as before. Moreover, in this thin limit,
$h$ completely disappears from the energy-momentum tensor and
the electric
flux. This is also true for Wess-Zumino term (\ref{wzterm}) since
$dT = \partial_wT\,dw$  kills the terms containing $h$ in
(\ref{dtfb}), which means that the constant magnetic field $h$
does not induce D2-brane charge.
>From the expression for the energy density in (\ref{10tew}), it
can be seen  that a BPS sum rule simlar to (\ref{bpssr})
holds and is independent of $h$.
Therefore, in the present case, one may say that  there is an
infinite degeneracy with respect to $h$ which is, different
from the previous case, not constrained to be zero in the BPS
limit. The BPS object is again a D4D2F1-composite with the
worldvolume direction $w$ replaced by $u$.

Since the range of the variable $w$ is from 0 to $4\pi R$, $\Sigma(w)$ now
reduces to sum of four delta functions rather than one in the thin
limit, i.e.,
\begin{equation} \label{thinw}
\Sigma(w) \longrightarrow \frac{\pi R {\cal T}_5}\alpha \sum_{n=0}^3
\delta(w-\pi R/2 - n\pi R), \qquad \beta\rightarrow 0.
\end{equation}
Due to this difference and the different pattern of degeneracy in $h$,
it might seem that the solution in this case describes a different object
from that of the previous case.
We will, however, show that the difference is actually
an artifact of the coordinates. It turns out that the two
configurations are identical, with the ways T$^2$ being embedded
in S$^3$ different.

First, we observe that, if the coordinate $w$ is exchanged with $u$,
the energy-momentum tensor (\ref{10tew}) and the electric flux
(\ref{fluxw}) are identical to (\ref{10te}) and (\ref{flux}) of
the previous case in the thin limit with $h=0$.
Now we consider the geometry of the solution. With $w = (n+\frac
12)\pi R$,
$n=0,\cdots,3$, the metric of S$^3$ reduces to
\begin{equation}
ds_{{\rm S}^{3}}=du^{2}
+ \sin^2 \left(\frac{u}{R}\right)\, dv^2 +\left[dw + \cos
\left(\frac{u}{R}\right)\, dv\right]^{2}
\longrightarrow du^2 + dv^2 ,
\end{equation}
which is locally the metric of T$^2$. The global topology is determined
by examining the range of coordinates. In the Appendix, we verify this
by finding an explicit orthogonal coordinate transformation connecting
the two solutions. Since the background fields are obviously invariant
under the transformation, the solution indeed describes the same object
as in the previous section embedded in a different direction in
S$^3$
(in the limit $\beta \rightarrow 0$).

\setcounter{equation}{0}
\section{Conclusion}
In this paper we studied the DBI type effective action of a
non-BPS D5-brane  in the asymptotic limit of the ten-dimensional
lift of the Salam-Sezgin vacuum.   In the limit the background
approaches R$^{1,4}\times$T$^2\times$S$^3$ and exact tachyon kink
solutions were found for the D5-brane on ${\rm R}^{1,2}\times {\rm
S}^3$. There are two classes of  solutions   both of which
describe BPS D4D2F1-composites on R$^2\times$T$^2$, D2 being
tubular and wrapped on an S$^1$ and F1 stretched along one of the
flat directions. In one class of the solutions, there is an
infinite degeneracy with respect to a constant magnetic field
along the direction of NS-NS field on the three-sphere.

Although the ten-dimensional embedding of the Salam-Sezgin vacuum
is a solution of type IIA supergravity, it needs to be seen
whether it is also an exact string background. One of the methods
to check this is to see whether the background survives
higher-order stringy correction terms in the low-energy type IIA
string effective action. However, the background at the
asymptotic limit may provide an exact string background,
considering that the local geometry is that of the near-horizon
limit of NS5-brane which is known to be exact at string tree
level~\cite{Callan:1991at}.

\section*{Acknowledgements}
This work was supported by Korea Research Foundation Grants
(KRF-2002-070-C00025 for C.K.) and is
the result of research activities (Astrophysical Research
Center for the Structure and Evolution of the Cosmos (ARCSEC))
supported by Korea Science $\&$ Engineering Foundation~(Y.K., H.K.,
and O.K.). C.K. and Y.K. thank the hospitality of KIAS where part
of the work has been done.

\appendix

\setcounter{equation}{0}
\section{Equivalence between the Two Types of Solutions in BPS Limit}

Here we demonstrate that under a suitable coordinate transformation
the solution $T=T(u)$ of (\ref{tub}) in section 3.1 becomes
the solution $T=T(w)$ of (\ref{tub9}) in section 3.2 in the thin limit.

The three-sphere with metric $ds_{\rm S^3}$ in (\ref{dst2}) can be
represented by Cartesian coordinates $\zeta_i$, $i=1,\cdots,4$ via
\begin{eqnarray}
\zeta_1 &=& \cos\left(\frac u{2R}\right)\,
            \cos\left(\frac{v+w}{2R}\right),\nonumber \\
\zeta_2 &=& \cos\left(\frac u{2R}\right)\,
            \sin\left(-\frac{v+w}{2R}\right),\nonumber \\
\zeta_3 &=& \sin\left(\frac u{2R}\right)\,
            \cos\left(\frac{v-w}{2R}\right),\nonumber \\
\zeta_4 &=& \sin\left(\frac u{2R}\right)\,
            \sin\left(\frac{v+w}{2R}\right),
\end{eqnarray}
so that $\sum_{i=1}^4 \zeta_i^2 = 1$. In these coordinates, the
solution
(\ref{tub}) describes T$^2$ embedded in S$^3$ as seen in section 3.1.
To visualize this solution, it is convenient to use the stereographic
projection of S$^3$ onto R$^3$ given by
\begin{equation}
w_i = \frac{\zeta_i}{1-\zeta_4},\qquad (i=1,2,3),
\end{equation}
where the point $(0,0,0,1)$ is chosen to be the ``northpole.'' Figure~1 shows
the torus (\ref{dst2}) embedded in S$^3$.

\begin{figure}
\centerline{\includegraphics[width=75mm]{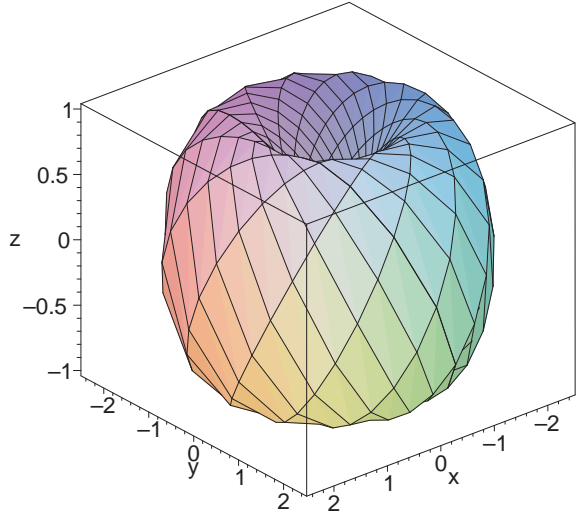}}
\caption{Stereographic projection of the solution $T=T(u)$ of section 3.1.}
\end{figure}

In this stereographic projection, the configuration obtained as the
thin limit of the solution (\ref{tub9}) is visualized as in Fig.~2.
Note that the figure is obtained after patching all contributions from
the four delta functions in (\ref{thinw}). At first, the shape of the surface
in Fig.~2 does not look like a torus. However, the spatial infinities
of R$^3$ are to be identified in the stereographic projection and it
is not difficult to see that the surface actually is a torus embedded in
S$^3$. Indeed, with the new coordinates
\begin{eqnarray}
\zeta'_1 &=& -\frac1{\sqrt2}(\zeta_2 + \zeta_4), \nonumber \\
\zeta'_2 &=& \frac1{\sqrt2}(\zeta_1 + \zeta_3), \nonumber \\
\zeta'_3 &=& \frac1{\sqrt2}(\zeta_2 - \zeta_4), \nonumber \\
\zeta'_4 &=& \frac1{\sqrt2}(\zeta_1 + \zeta_3),
\end{eqnarray}
and the corresponding stereographic projection, the surface in Fig.~2 is
precisely transformed to the torus of Fig.~1.

\begin{figure}
\centerline{\includegraphics[width=75mm]{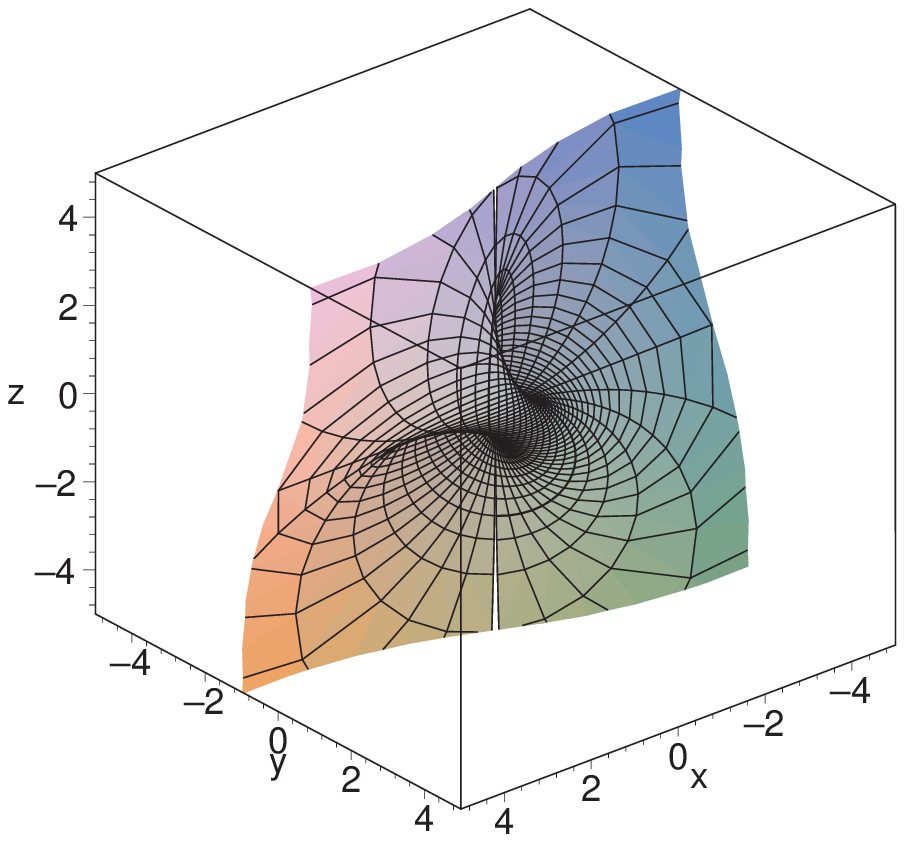}}
\caption{Stereographic projection of the solution $T=T(w)$ of section 3.2.}
\end{figure}

\end{document}